\documentstyle[twocolumn,prl,aps]{revtex}

\begin{document}
\draft
\preprint{}
\title{Critical Properties of Spectral Functions \\
for the 1D Anisotropic t-J Models with an Energy Gap}
\author{ Tatsuya Fujii, Yasumasa Tsukamoto and Norio Kawakami } 
\address{Department of Applied Physics,
Osaka University, Suita, Osaka 565, Japan} 
\date{June 7, 1997}
\maketitle
\begin{abstract}
We exactly calculate the momentum-dependent critical 
exponents for spectral functions in the one-dimensional 
anisotropic {\it t-J} 
models with a gap either in the spin or charge excitation 
spectrum. Our approach is based on the Bethe ansatz technique 
combined with finite-size scaling 
techniques in conformal field theory. It is found that the spectral 
functions show a power-law singularity, 
which occurs at frequencies determined by the 
dispersion of a massive spin (or charge) excitation.
We discuss how the nontrivial 
contribution of a massive excitation controls the singular behavior 
in optical response functions. 
\end{abstract}
The dynamical spectral properties of one-dimensional (1D) highly 
correlated electron systems have been intensively studied.
For example, there are  a number of 
recent experiments  on angular resolved 
photoemission,\cite{ex1,ex2,ex3,ex4,ex5} which have revealed
some striking properties inherent in 1D systems, such as
the spin-charge separation.\cite{ex5}  Also, 
extensive theoretical studies on spectral functions
have been done numerically \cite{preuss,haas,maekawa,loren,mila,shiba} and 
analytically. \cite{meden,voit2,nagaosa} 

A remarkable point in the recent photoemission 
experiment\cite{ex2} for the insulator compound SrCuO$_{2}$
is that it has successfully explored characteristic 
properties of {\it a single hole} doped in the 1D Mott insulator.
In such insulating systems, only the spin sector is
massless (Tomonaga-Luttinger (TL) liquid), 
while the charge sector is massive.
This type of 1D electron system is sometimes referred to  
as the Luther-Emery class, for which either spin or charge
excitation is massive. It was naively believed that
the one-particle Green function 
for such systems with massive excitations
would show exponential decay instead of the power-law singularity.
However, Sorella and Parola,\cite{S&P1,S&P2} and Voit\cite{voit} 
theoretically predicted  that the dynamical
spectral functions for these systems 
still show the  power-law singularity at threshold
energies, characterizing 
the critical properties of a doped hole in the 1D Mott insulator.
In particular, Sorella and Parola\cite{S&P2}
have demonstrated the above point by using the 
exact solution of the supersymmetric {\it t-J} model and  conformal 
field theory (CFT).

Motivated by the above investigations, we study the spectral 
properties of 1D highly correlated electron systems with 
a gap either in the spin or charge excitation spectrum.
For this purpose, we consider two types of instructive models for  
1D correlated electron systems: the first model is 
the integrable anisotropic {\it t-J} model\cite{bariev} 
away from half filling, for which
the charge (spin) sector is massless (massive).
The other one is the
hole-doped XXZ Heisenberg spin chain, which is realized by
the anisotropic {\it t-J} model in the limit of half filling
with the condition $J{\rightarrow}0$.
We explore this model as a typical system which has  
massless spin (massive charge) excitations.
We exactly calculate the critical exponents 
of spectral functions for both of these models, by exploiting 
finite-size scaling ideas in CFT.  
We discuss the critical properties of 
the optical response functions in terms of the momentum-dependent
critical exponents obtained in this paper.

%


Let us start with the 1D anisotropic 
{\it t-J} model, which possesses
a gap for spin excitations. This model is a 
deformed supersymmetric {\it t-J} model
which was exactly solved by Bariev,\cite{bariev}
\begin{eqnarray}
{\cal H}_{1} &=& -\sum_{i,\sigma} {\cal P}
( c^{\dagger}_{i,\sigma}c_{i+1,\sigma}
                         + c^{\dagger}_{i+1,\sigma}c_{i,\sigma} ) {\cal P } \cr
             &+& \sum_{i} (c^{\dagger}_{i,\downarrow}c_{i,\uparrow}c_{i+1,\uparrow}^{\dagger}c_{i+1,\downarrow} 
             +c^{\dagger}_{i,\uparrow}c_{i,\downarrow}c^{\dagger}_{i+1,\downarrow}c_{i+1,\uparrow}  \cr
       && {\hskip7mm}   - e^{-\eta}n_{i,\downarrow}n_{i+1,\uparrow} 
                        - e^{\eta}n_{i,\uparrow}n_{i+1,\downarrow} ),
\label{eqn:1Danisotropic}
\end{eqnarray}
where the operator 
${\cal P} = {\Pi}^{N}_{j=1}(1-n_{j,\uparrow}n_{j,\downarrow})$ 
projects out the Hilbert space without double occupancy
at all sites, and $\eta$ is a deformation 
parameter which parametrizes the anisotropy of the 
interaction. In the isotropic limit (${\eta}{\rightarrow}0$), 
the Hamiltonian (\ref{eqn:1Danisotropic}) 
reduces to the conventional supersymmetric {\it t-J} model 
which was exactly solved,\cite{southern,sh,bares1,bares2}$\hskip -1mm$ and its
conformal properties were clarified.\cite{kawakami2} 

It is known that the spin excitation for (\ref{eqn:1Danisotropic})
is always massive, while the charge excitation is massless away from  
half filling.\cite{bariev,khomoto&sato}  Hence,this model
allows us to discuss the spectral properties of 
the system with the massless charge and
massive spin excitations.
We wish to emphasize here that although
the following analysis is performed for the above specific model, the 
obtained results can be generally applied to 1D correlated
electron  models with the
spin gap, such as the attractive Hubbard model.

In the following, we derive the one-particle Green function
in the low-energy regime and the corresponding critical exponents.
To this end, we derive the relevant Bethe equations which specify 
excitations when one electron is removed from the system.
These equations can be straightforwardly deduced from those obtained 
by Bariev.\cite{bariev}  The result reads
\begin{eqnarray}
 \left(  \frac{ {\sin} ( {\Lambda}_{\alpha} + {\rm i} {\eta} ) } 
{{\sin} ( {\Lambda}_{\alpha} - {\rm i} {\eta} ) }  \right)^{N}  
 & = &
- \frac{ {\sin} ( {\Lambda}_{\alpha} - {\lambda}+ {\rm i} {\eta}/{2} ) } 
{ {\sin} ( {\Lambda}_{\alpha} -{\lambda} - {\rm i} {\eta}/{2} ) }  \cr
& & \times \prod^{M}_{\beta = 1}
\frac{ {\sin} ( {\Lambda}_{\alpha} - {\Lambda}_{\beta}+ {\rm i} {\eta} ) }
 { {\sin} ( {\Lambda}_{\alpha} -{\Lambda}_{\beta} - {\rm i} {\eta} ) }, 
\label{eqn:bethe}
\end{eqnarray}
where $N$ ($M$) is the number of sites (down-spin electrons), 
and  $\Lambda_{\alpha}$ 
($\alpha=1,2,\dots,M$) are charge rapidities which classify
collective charge excitations.  Note that  
a spin rapidity $\lambda$ also appears in the above
equation, which characterizes a  massive spin excitation induced
by the  injected hole.
Although massive spin excitations do not directly enter in the low-energy 
physics, they control the critical 
behavior of massless charge excitations via 
the nontrivial coupling between charge and spin sectors represented 
by (\ref{eqn:bethe}), as will be shown momentarily.
%

In order to apply the sophisticated methods developed in  CFT,\cite{bpz} 
we now analyze the finite-size corrections to the ground-state 
energy and excitation energy
including the effect of the massive spin sector.\cite{cardy,affleck}
 This manipulation  enables us to exactly calculate 
the critical exponents for spectral  functions  in the 
low energy regime.
The calculation is straightforwardly performed 
in a standard manner,\cite{woy,kawakami,fk}
to yield the following expressions, which are typical for 
TL liquids classified by $c=1$ CFT,
\begin{eqnarray}
E_{0} & = & N{\epsilon}_{0} - \frac{ {\pi} v_{c} }{6N}, \cr
E-E_{0} & = & {\Delta} {\epsilon}_{s} + {\Delta} {\epsilon}_{c} =
            e_{s}({\lambda}) + { \frac{ 2 {\pi} v_{c} }{N} } x_{c},
\label{eqn:size correction}
\end{eqnarray}
where $N{\epsilon}_{0}$ is the ground-state energy in the
thermodynamic limit.
The $N$-independent term $\Delta \epsilon_s 
=e_{s}$ (so-called surface energy) 
gives an excitation energy for the massive spin sector. 
It is  clearly seen that the spin excitation  
$ {\Delta} {\epsilon}_{s}$ is decoupled from the  charge excitation
$ {\Delta} {\epsilon}_{c}$. 
According to the finite-size scaling idea in CFT,
the scaling dimension for the charge sector is read 
from the universal $1/N$ corrections to the 
excitation energy $ {\Delta} {\epsilon}_{c}$, which is given as
\begin{eqnarray}
x_{c} = \frac{1}{ 4 \xi_{c}^{2} } ( \Delta M_{c} + n^{(s)})^{2} + 
\xi_{c}^{2} ( \Delta D_{c} + d^{(s)})^{2},
\label{cscaling}
\end{eqnarray}
where ${\Delta} D_{c}$ and $\Delta M_{c}$ are the quantum numbers
 that label current excitations and charged excitations,
which are subjected to the selection rule
for fermions,  ${\Delta} D_{c} = -{\Delta} M_{c}/2$.
The quantity $\xi_c=\xi_c(v_0)$, often  referred to  as the dressed 
charge, is given as\cite{bariev}
\begin{eqnarray}
\xi_{c} (v) = 
      1  -   \left( \int^{-v_{0}}_{-\pi} + \int^{\pi}_{v_{0}}\right)
          \frac{1}{2\pi}  \Theta^{'}(v-v^{'},\eta)  
               \xi_{c} (v^{'}){\rm d}v^{'}, \cr
\end{eqnarray}
where
\begin{eqnarray}
     \Theta(v,\eta)  =  2\tan^{-1} (\coth(\eta)\tan(v/2)).
\end{eqnarray}
Note that the dressed charge features the U(1) 
critical line of $c=1$ CFT
when we change the strength of interaction or the density of electrons.
Here, we stress that the scaling dimension  $x_{c}$ 
for the charge sector completely
determines the critical behavior of the spectral functions.
Note, however, that  two key quantities $n^{(s)}$ and $d^{(s)}$, 
which contain the effect of the massive spin sector, 
are introduced in (\ref{cscaling})
\begin{eqnarray}
n^{(s)} &=& \left( \int^{-v_{0}}_{-\pi} + 
\int^{\pi}_{v_{0}}  \right) \sigma_{s}(v){\rm d}v, \cr
d^{(s)} &=& \frac{1}{2} ( z_{s}(\pi)+z_{s}(-\pi) )   \cr
     && + \frac{1}{2} \left( \int^{-v_{0}}_{-\pi} 
- \int^{\pi}_{v_{0}} \right)\sigma_{s}(v){\rm d}v, 
\end{eqnarray}
where
\begin{eqnarray}
z_{s}(v)  &=&   \Theta(v-2{\lambda},\eta/2)    \cr
      && - \left( \int^{-v_{0}}_{-\pi} + \int^{\pi}_{v_{0}}  \right)
             \Theta(v-v^{'},\eta)  \sigma_{s} (v^{'}){\rm d}v^{'}, \cr
{\sigma}_{s}(v)  &=&   \frac{1}{2\pi} \Theta^{'}(v-2{\lambda},\eta/2) \cr
    &&   - \left( \int^{-v_{0}}_{-\pi} + \int^{\pi}_{v_{0}}\right)
        \frac{1}{2\pi}  \Theta^{'}(v-v^{'},\eta)  
\sigma_{s} (v^{'}){\rm d}v^{'}. 
\nonumber
\end{eqnarray}
These quantities are alternatively represented in terms
of the phase shifts $\delta_L$ and 
$\delta_R$ at the left and right Fermi points in the 
massless charge sector:  $n^{(s)}=\delta_R+\delta_L$
 $d^{(s)}=\delta_R-\delta_L$.
Although the scaling dimension  
$x_{c}$ reflects $c=1$  CFT  for the massless sector,
the massive spin sector also contributes 
to the scaling dimension  via $n^{(s)}$ and $d^{(s)}$, 
as seen  in (\ref{cscaling}). 
Note that the eq.(\ref{cscaling}) for scaling dimensions 
is typical for   {\it shifted} $c=1$ CFT,
whose fixed point is different from that of
the static impurity problem, as pointed out by Sorella 
and Parola.\cite{S&P2,tsukamoto}

We are now ready to investigate the critical properties of the 
one-particle Green function
 $ G(x,t)$. According to
the spin-charge separation in elementary 
excitations for 1D systems, it is given by,
\begin{eqnarray}
   G(x,t) &=& {\rm i} \sum_{\lambda} {\theta}(-t) {\rm e}^{{\rm i}{\Delta}{\epsilon}_{s}t   
                                                    -{\rm i}{\Delta}p_{s}x}  \cr
         & &{\times}< {\Psi}_{0} |{\hat {\psi}}^{\dagger}(0) |{\lambda}>
                <{\lambda} | {\hat {\psi}}(0) | {\Psi}_{0} >
                   {\rm e}^{{\rm i}{\Delta}{\epsilon}_{c}t - {\rm i}{\Delta}p_{c}x}  \cr
{\baselineskip5mm} \cr
          &=& {\hskip1.5mm} \sum_{\lambda}G_{s}(x,t) {\times} G_{c}(x,t).
\label{eqn:G}
\end{eqnarray}
As is well known, the Green function $G_s$ for the massive 
spin sector gives the exponential decay, {\it e.g.}, in the long-distance
behavior of the equal-time Green function $G(x,t=0)$.
However, this is not true for the dynamical Green function 
$G(k,\omega)$ around $\omega=\omega_s$, for which the massive spin 
excitation $w_s$ determines the shift of the energy, whereas
the massless charge excitation still causes the 
infrared singularity, resulting in the 
power-law behavior around $\omega=\omega_s$.
In order to clarify how such a power-law singularity appears, 
we first concentrate on the low-energy behavior of the Green 
function $G_{c}$ in the massless charge sector.
Exploiting  finite-size scaling techniques in CFT, we can write down 
its asymptotic form as 
\begin{eqnarray}
G_{c}(x,t) {\sim} \frac{ {\rm e}^{{\rm i}Q_{c}x} }{ (x-{\rm i}v_{c}t)^{ 2{\Delta}^{+} }
                   (x+{\rm i}v_{c}t)^{ 2{\Delta}^{-} } },
\end{eqnarray}
where $v_c$ is the velocity of charge excitations and
$Q_{c} = 2Q_{F} ( {\Delta} D_{c} + d^{(s)})$.  Here  
${\Delta}^{\pm}$ are conformal dimensions which 
are related to the  scaling dimension 
$x_{c}$, 
\begin{eqnarray}
x_{c} = {\Delta}^{+} + {\Delta}^{-}.
\label{eqn:relation}
\end{eqnarray}
To get the critical exponent for the one-particle Green function,
we have to set  ${\Delta} M_{c}=-1$ and ${\Delta} D_{c}=1/2$ in 
(\ref{cscaling}).

The spectral function $A(k,{\omega})$ is now given 
via the Fourier transformation as 
\begin{eqnarray}
A(k,{\omega}) = {\frac{1}{\pi}} Im G(k,{\omega})
              {\sim} ({\omega} - {\omega}_{s}(k-Q_{c}))^{X(k)},
\end{eqnarray}
with the critical exponent $X(k)=2x_{c}-1$.
It is seen from this formula
 that the most relevant singularity in the spectral function 
occurs at frequencies determined by the massive 
spin spectrum ${\omega}_{s}(k-Q_{c})$, 
and the corresponding critical exponent $X(k)$ is dependent 
on the momentum $k$ of an injected hole.  
Note that the quantities which  
give rise to the momentum dependence of the critical exponent 
are the phase shifts $n^{(s)}$ and $d^{(s)}$.

The momentum dependence of a critical exponent $X(k)$ is 
shown in Fig. 1. Here the 
momentum is plotted in the unit of inverse 
lattice spacing $1/a$.
It is clear that $X(k)$ has values 
with either a positive  or a negative sign, which respectively 
results in convergence or divergence power-law behavior around
 the threshold energy  ${\omega}_{s}(k-Q_{c})$.
  Therefore, we can say that the spectral intensity around
the threshold would be
strongly momentum dependent. In particular, for
the region exhibiting convergence power-law behavior, i.e.
for $X(k)>0$, the edge singularity is completely smeared, and 
may not be observed clearly in experiments. 
We note that  the plateaulike structures 
in part of the momentum region in Fig. 1 for the case of 
small $\eta$ (close to isotropic case) reflect the fact
that $n^{(s)}$ is weakly dependent on the momentum 
for the case of  $\eta \sim 0$. 

%

We now proceed to another interesting anisotropic {\it t-J} model, 
\begin{eqnarray}
{\cal H}_{2} =      &-&t {\hskip2mm} {\cal P} \{ {\hskip1mm} \sum_{i,\sigma} 
                   ( c^{\dagger}_{i,\sigma}c_{i+1,\sigma} 
                     + c^{\dagger}_{i+1,\sigma}c_{i,\sigma} ) {\hskip1mm} \} {\cal P } \cr
                    &-&J  \sum_{i} { \{ } {\sigma}^{x}_{i}{\sigma}^{x}_{i+1} 
                                        + {\sigma}^{y}_{i}{\sigma}^{y}_{i+1}
                          + \cos(2\eta) ( {\sigma}^{z}_{i}{\sigma}^{z}_{i+1} -1 )  \cr
                        &&+ H({\sigma}^{z}_{i}-1)    { \} },
\label{eqn:xxz} 
\end{eqnarray} 
where ${\sigma}$ are Pauli matrices, $\eta$ is a non-negative
quantity which 
parametrizes the anisotropy, and $H$ is an applied magnetic field. 
This model is realized  by doping holes  
into the integrable 1D XXZ spin model.\cite{bogo}  If the condition
$J/t \ll 1$ is satisfied, we can perform the exact 
evaluation of the critical exponents based on the idea
of CFT. In the following, we wish to deal with
the case where a single hole is doped into the above insulating
XXZ spin chain, for which the spin excitation still remains massless.
Note that  the present system belongs to
the same universality class as
the repulsive Hubbard model at half filling, and the
results obtained below should be complementary to
those of Sorella and Parola for the supersymmetric case.\cite{S&P2}

We now investigate the critical behavior of the 
spectral function at half filling 
in the limit $J{\rightarrow}0$.  Repeating the same calculation 
as for the previous model (\ref{eqn:1Danisotropic}), 
we immediately end up with the following scaling dimension for  
spin excitations,
\begin{eqnarray}
x_{s} = \frac{1}{ 4 \xi_{s}^{2} } ( \Delta M_{s} + n^{(c)})^{2}
 + \xi_{s}^{2} ( \Delta D_{s} + d^{(c)})^{2},
\label{eqn:xs}
\end{eqnarray}
where the quantum numbers have been chosen
as $\Delta M_{s}=-1$ and $\Delta D_{s}=-1/2$ 
to satisfy the fermion selection rule $\Delta D_{s}=\Delta M_{s}/2$.
The dressed charge $\xi_{s}=\xi_{s}(\Lambda_0)$, is given by
\begin{eqnarray}
   \xi_{s} (\Lambda) = 1  +   \int^{\Lambda_0}_{-\Lambda_0} 
                             \frac{1}{2\pi} K (\Lambda - \Lambda^{'},\eta)  
                             \xi_{s} (\Lambda^{'}){\rm d} \Lambda^{'}, \cr
\end{eqnarray}
where
\begin{eqnarray}
  K (\Lambda , \eta) = \frac{\sin 4\eta}{{\rm sinh}(\Lambda + {\rm i}2\eta)
{\rm sinh}(\Lambda - {\rm i}2\eta)}.
\end{eqnarray}
The charge contribution to the massless spin sector
is explicitly written down in terms of two phase shifts
\begin{eqnarray}
n^{(c)} = \frac{1}{2}(m+1), {\hskip5mm} d^{(c)} = 0,
\end{eqnarray}
where $m$ is the magnetization of the system in the 
unit of $g\mu_B=1$. Note that 
the spin and charge sectors  in the present model
exchange their role
compared with  the {\it t-J} model (\ref{eqn:1Danisotropic}).
It turns out that in the limit $J{\rightarrow}0$,  
$n^{(c)}$ and $d^{(c)}$ do not depend 
on the momentum of an injected hole, in contrast
to the model (\ref{eqn:1Danisotropic}).
The critical behavior of the spectral function $A(k,{\omega})$
is determined by the above scaling dimension,
\begin{eqnarray}
A(k,{\omega}) = {\frac{1}{\pi}} Im G(k,{\omega})
              {\sim} ({\omega} - {\omega}_{c}(k-Q_{s}))^X,
\label{eqn:spectral2}
\end{eqnarray}
where $Q_{s}=2Q_{F}(\Delta D_{s} + d^{(c)})$ 
and ${\omega}_{c}(k)=-2\cos(k)$ is the excitation spectrum for the 
massive charge sector.  

The computed critical exponents $X$ are shown in Fig. 2
as a function of the magnetization.
For the completely polarized case ($m=1$),
the dressed charge $\xi_{s}$ equals to unity as 
in the noninteracting case. Therefore, the critical exponent 
takes the same constant value, $-1/2$, even if the
anisotropy parameter $\eta$ is changed. 
In the limit $\eta \rightarrow \pi/4$ 
the XXZ model reduces 
to the XY model, which is nothing but a free fermion model. 
Although  it is naively expected that the critical exponent 
for such a free system is trivial,  
the presence of two phase shifts $n^{(c)}$ and $d^{(c)}$ 
result in the nontrivial exponents, as is the 
case for the ordinary X-ray singularity problem.
In the opposite limit, $\eta \rightarrow \pi/2$, the XXZ model reduces 
to the XXX model. The resulting critical exponent $-1/2$ at $m=0$ 
exactly coincides with the one obtained 
by Sorella and Parola,\cite{S&P1} and Voit.\cite{voit}

In summary, we have studied the critical properties of spectral 
functions for the 1D anisotropic {\it t-J} models
combining the Bethe ansatz technique 
with finite-size scaling methods in CFT. 
In particular, we have  focused on the models possessing
an excitation gap either in the spin or charge 
spectrum by exploiting the two anisotropic {\it t-J} models.
By investigating  critical properties  exactly,
we have seen that spectral functions show power-law 
behavior around the threshold energies,
as found by Sorella and Parola for the isotropic 
supersymmetric model. 
The singularity in spectral functions occurs 
at frequencies determined by the dispersion of a massive 
excitation, reflecting the nontrivial interaction 
between two elementary excitations.
It is interesting to extend the present analysis 
to the case with orbital degeneracy,
which would  provide useful information 
about optical response functions for some Mn oxides, which have been 
investigated intensively  in recent years. 

We would like to thank S. Fujimoto, T. Fukui and  S. Kumada 
for their valuable discussions.
This work was partly supported by a Grant-in-Aid from the Ministry
of Education, Science, Sport and Culture, Japan.

%

%
\begin{figure}[h]
\vspace*{1cm}
\caption{Momentum-dependent critical exponents $X(k)$ for the 
spectral function in the 1D anisotropic {\it t-J}
model(1) 
for which the spin 
sector is massive and the charge sector is massless.}
\end{figure}
\begin{figure}[h]
\vspace*{1cm}
\caption{Critical exponents $X$ for the spectral function
as a function of the magnetization $m$. The model is the hole-doped 
Heisenberg XXZ spin chain in the limit of 
$J/t \ll 1$, for which the spin sector is massless and
the charge sector is massive.
}
\end{figure}
\end{document}